\documentclass[epj]{svjour}  
\usepackage{epsfig}
\begin{document}
\title{Transport and Coulomb drag for two interacting carbon nanotubes} 
\author{A.~Komnik and R.~Egger}
\institute{
Fakult\"at f\"ur Physik, Albert-Ludwigs-Universit\"at,
D-79104 Freiburg, Germany}
\date{Date: \today}
\abstract{
We study nonlinear transport for two coupled one-dimensional quantum
wires or carbon nanotubes described by Luttinger liquid theory. 
Transport properties are shown to crucially depend on the contact 
length $L_c$. For a special
interaction strength, the problem can be solved analytically for
arbitrary $L_c$.  For point-like contacts and 
strong interactions, a qualitatively different picture
compared to a Fermi liquid emerges, characterized by zero-bias
anomalies and strong dependence on the applied cross voltage.
In addition, pronounced Coulomb drag phenomena are important 
for extended contacts.  
}


\maketitle

\section{Introduction}
\label{secI}

Transport in interacting one-dimensional (1D) quantum wires (QW) has attracted
ever-increasing attention over the past decade.  This interest was sparked
mainly by the discovery of novel 1D materials besides standard
(semiconductor or organic chain molecule) systems,
such as edge states in fractional quantum Hall
bars or carbon nanotubes.  Furthermore, 1D QWs are predicted to behave as a
Luttinger liquid (LL) due to electron-electron interactions \cite{gogolin}.
Recent nonlinear transport experiments \cite{lltube1}
for individual nanotubes 
have indeed demonstrated impressive
agreement with the LL theory of nanotubes \cite{lltube2}.
In these experiments, transport was limited either by the 
contact resistance to the leads,
or by a tunnel junction (``topological kink'') within the nanotube.
In both cases, the observed power laws in the (nonlinear) conductance have
allowed for a consistent 
explanation in terms of LL theory.

Different transport experiments built up of at least two 
nanotubes can reveal even more dramatic deviations from
Fermi liquid transport.  The theoretical predictions of 
Ref.~\cite{xll} for crossed nanotubes (which are 
coupled in a pointlike way) were recently
observed experimentally by Kim {\sl et al.}~\cite{kim}.
For longer contacts between the nanotubes, 
Coulomb drag \cite{rojo} is expected to play an important role
in addition to the crossed nanotube scenario. 
Coulomb drag can be very pronounced in one dimension 
and leads to quite rich physics.
In this paper, we study in detail two 
nanotubes arranged parallel to or crossing each other,
and briefly discuss more complex setups.   Notably, such experiments
are feasible using present-day technology
\cite{kim,fuhrer}.

\begin{figure}
\begin{center}
\epsfxsize=1.0\columnwidth
\epsffile{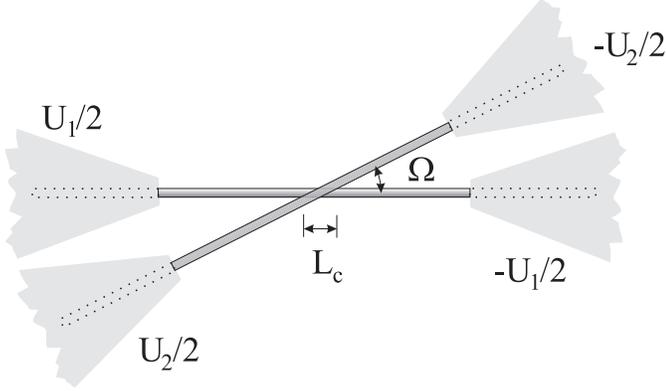}
\caption[]{\label{fig1} 
Two crossed nanotubes in contact to reservoirs for
applied voltage $U_1$ and $U_2$, respectively (schematic view). 
Different crossing angles $\Omega$ imply 
different effective contact lengths $L_c$.
}
\end{center}
\end{figure}

The main part of the paper focuses on the schematic geometry  
shown in Figure \ref{fig1}, where transport through two
clean (ballistic) nanotubes biased by voltages $U_{1,2}$ is studied.  The
nanotubes are brought to contact by crossing them under an angle $\Omega$.
By varying this angle, the effective contact length $L_c$ 
can be changed.  We shall address the crossover from a 
point-like crossing, where $L_c\approx a$ with the lattice spacing $a$, 
to an extended coupling, where $a\ll L_c \leq L$ with the tube length $L$.
Besides our previous paper \cite{xll},
theoretical predictions for transport in such a geometry have been
given by other authors
\cite{flensberg,drag,klesse,ponav}.  Their results
were largely obtained in the  linear regime,
or focus on either very small or very large $L_c$.
In this paper, we cover the full crossover from a point-like
to an extended contact, and explicitly compute
{\sl nonlinear} current-voltage relations.  
For simplicity, we consider the same interaction 
strength parameter $g$ and Fermi velocity $v_F$ 
for both QWs. The LL parameter $g$ equals 
unity for a Fermi gas, and becomes smaller for strong repulsive interactions.
Provided one works on an insulating substrate,
the LL parameter has only a weak logarithmic
dependence on the tube length $L$,
and the experimentally observed value $g\approx 0.25$ 
\cite{lltube1,lltube2,kim}
indicates strong non-Fermi liquid behavior.  

For repulsive interactions ($g<1$),
the most important coupling between the two nanotubes 
is of electrostatic origin
\cite{xll}.  It is responsible for the most relevant operator
under the renormalization group (RG), and can cause perfect
Coulomb drag at low temperatures.
In addition, as indicated by the data of Ref.~\cite{fuhrer},
electron tunneling between the QWs can be 
important. Provided one has sufficiently strong
interactions, tunneling into a LL is {\sl irrelevant}\
in the RG sense.  In that case, one can treat it perturbatively, at least
for not exceedingly large tunneling amplitude.  Notice that
otherwise our assumption of clean wires breaks down in any case,
since good mechanical contact of the QWs indicates impurity formation  
within each QW.  

The outline of our paper is as follows. In Sec.~\ref{secII}, we briefly
review the LL concept and discuss how the theory can incorporate 
adiabatically coupled voltage reservoirs
in terms of Sommerfeld-like radiative 
boundary conditions \cite{egger96}.
Sec.~\ref{secIII} presents a detailed discussion of transport
for the setup in Fig.~\ref{fig1} under the assumption of
negligible tunneling. In particular, for the special interaction
strength $g=1/2$, the full temperature-dependent transport
problem is solved for arbitrary contact length $L_c$.
The role of adiabaticity of the contacts to the reservoirs for the
described effects is addressed in Sec.~\ref{secX}, where we 
study weakly contacted reservoirs
in the tunneling limit.
In Sec.~\ref{secIV}, effects of inter-wire tunneling 
are discussed, and in Sec.~\ref{secV} some conclusions
and possible applications are outlined.
 
For clarity, we focus on spinless single-channel QWs.  
The modifications arising for 
spin-1/2 electrons or carbon nanotubes are then straightforward.
In addition, the relevant energy scale is supposed to exceed
$v_F/L$, so that we can effectively put $L\to \infty$.
Below we put $\hbar=e=k_B=1$ and $v=v_F/g=1$. 
 
\section{Model}
\label{secII}

The LL concept for 1D metals is 
most transparent in the bosonization representation \cite{gogolin}. 
The electron field operator is written as
superposition of left- and right-moving ($p=R/L=\pm$) 
fermions $\psi_{p\alpha}(x)$  for QW  $\alpha=1,2$.
The latter are expressed in terms of canonically conjugate
bosonic fields $\theta_\alpha(x)$ and $\phi_\alpha(x)$, 
\begin{eqnarray} \label{psi}
 \psi_{p\alpha}(x) &\sim& (2\pi a)^{-1/2}
 \exp [ -i p k_{F \alpha} x] \\ \nonumber
&\times& \exp[ - i p (\pi g)^{1/2} \theta_\alpha(x) -i (\pi/g)^{1/2}
 \phi_\alpha(x) ] \;,
\end{eqnarray}
where $a$ is a lattice constant.
The Fermi momentum $k_{F\alpha}$ in the two QWs can be 
made different by uniformly shifting the chemical potentials 
of both reservoirs attached to one wire.  The $k_{F\alpha}$ 
are defined such that the chemical potentials are $\pm U_{1,2}/2$
as indicated in Figure~\ref{fig1}.
Using Eq.~(\ref{psi}), the density operator $\rho_\alpha(x)$ is
\begin{equation} \label{density}
\rho_\alpha= (g/\pi)^{1/2} \partial_x \theta_\alpha
+ \frac{k_{F \alpha}}{\pi}  \sin \left[ 2 k_{F \alpha} 
 x + \sqrt{4 \pi g} \,\theta_\alpha \right] \;.
\end{equation}
The uncoupled clean QWs correspond to the standard LL Hamiltonian, 
\begin{equation} \label{H}
H_0 = \sum_\alpha \frac12 \int dx \, \left[
(\partial_x \phi_\alpha)^2 + (\partial_x \theta_\alpha)^2 \right]  \; .
\end{equation}

Let us next address how the coupling of each wire
to the voltage reservoirs can be taken into account.
For adiabatic coupling, the electron densities near the end of the QW
obey the radiative boundary conditions \cite{egger96}
\begin{equation}\label{bc}
\frac{g^{-2}\pm 1}{2}\rho_{R\alpha}(\mp L/2) + 
\frac{g^{-2}\mp 1}{2}\rho_{L\alpha}(\mp L/2)
= \pm \frac{U_\alpha}{4\pi g } \;,
\end{equation}
where the $p=R/L=\pm$ moving densities in wire $\alpha$ are 
\begin{equation} \label{densbos}
\rho_{p\alpha}(x) = (4\pi g)^{-1/2} \partial_x [ g\theta_\alpha
 + p\phi_\alpha] \;.
\end{equation}
These boundary conditions need to be enforced for the stationary expectation
values of the densities $\rho_{p\alpha}$ in the QW near
the respective contact to the leads.  They hold for arbitrary 
impurity scattering within each wire and therefore also in
the presence of coupling between the two wires.

In the absence of inter-wire tunneling, the charge current 
\begin{equation}\label{currbos}
I_\alpha(x) = e(g/\pi)^{1/2} \partial_t \theta_\alpha(x) 
\end{equation} 
is conserved and independent of $x$.  In the presence of
tunneling, however, we need to distinguish $I_\alpha(x<-L_c/2)$
and $I_\alpha(x>L_c/2)$. Postponing the discussion
of inter-wire tunneling to Sec.~\ref{secIV},
the conductance of each (impurity-free) QW is thus $G_0=e^2/h$ in the
absence of {\sl electrostatic} inter-wire coupling.
The latter then implies a reduction of the conductance 
\begin{equation}\label{cond1}
G_\alpha = I_\alpha/ U_\alpha \;,
\end{equation}
since the transport-carrying density waves drag each other, 
which makes them  ``heavier.'' For low energy scales, 
the electrostatic coupling can be expressed as a {\sl local}\
product of densities in the two wires.
To justify the locality of inter-wire interactions, one
can employ the same reasoning as for the locality of the 
intra-wire interaction \cite{gogolin}.
Neglecting momentum-non-conserving terms, we obtain
\begin{eqnarray} \label{elop}
 H_1  & =& V_1 \int dx \, \zeta(x)  \rho_1(x) \rho_2(x) \\ \nonumber
 &=&  V_{1a} \int dx  \zeta(x) \partial_x
 \theta_1 \partial_x \theta_2 \\  \nonumber
 &+&  V_{1b} \int dx \zeta(x)  \sin \left[2 k_{F 1} x + \sqrt{4 \pi
 g} \,\theta_{1}(x) \right] \\ &\times&  \nonumber
 \sin \left[2 k_{F 2} x + \sqrt{4 \pi
 g} \,\theta_{2}(x) \right] \; ,
\end{eqnarray}
where $V_{1a}=g V_1 /\pi$ and
$V_{1b}=V_1 k_{F1} k_{F2}/\pi^2$.  
The function $\zeta(x)$ specifies the spatial dependence
of the inter-wire coupling, and for practical purposes,
we consider $\zeta(x)=1$ for $|x|\leq L_c/2$ and 
zero otherwise,
\begin{equation}\label{contact}
\zeta(x)= [\Theta(x+L_c/2)-\Theta(x-L_c/2)] \;.
\end{equation}
In addition, {\sl tunneling}\ leads to 
\begin{equation}
 H_2 =  V_2 \int  dx  \zeta(x)  \sum_{pp'} \psi_{p1}^\dagger(x)
\psi_{p'2}^{}(x) + {\rm h.c.}\;,
\end{equation}
describing electron transfer between the QWs.
The bo\-so\-nized form of $H_2$ can be found in Ref.~\cite{xll}.
We shall turn to a discussion of tunneling in Sec.~\ref{secIV}. 
Josephson-type couplings \cite{starykh}
are irrelevant for $g<1$ and
hence will be ignored in the following. 

\section{Electrostatically coupled nanotubes}
\label{secIII}

In this section, we assume that tunneling can be neglected,
and consider a system described by the Hamiltonian $H=H_0+H_1$
under the boundary conditions (\ref{bc}).
Below we separately discuss three cases, namely
 (a) a strictly local contact with $L_c\to 0$, (b)
a short but finite contact length $L_c\approx a$, and
(c) for the special
interaction strength $g=1/2$, we present an analytical 
solution valid for arbitrarily long contacts.
Of particular interest is the 
{\sl differential conductance matrix}, 
\begin{equation}\label{tcond}
G_{\alpha\alpha'} = \partial I_\alpha/\partial U_{\alpha'} \;. 
\end{equation}
The off-diagonal conductance $G_{12}$ (or $G_{21}$),
the so-called ``transconductance'',
is the appropriate quantity measuring Coulomb drag \cite{rojo}.
For $U_1=U_2=0$, the matrix $G_{\alpha\alpha'}$ describes
the linear conductances.  Generally,
the diagonal conductance $G_{\alpha \alpha}$ and the conductance $G_\alpha$
defined in Eq.~(\ref{cond1}) show qualitatively the same behavior.

\subsection{Strictly local contact}
\label{secIIIa}

Let us start with an ideal point-like contact, where from
Eq.~(\ref{contact}), we get for $L_c\to0$ the result
$\zeta(x)=L_c \delta(x)$.
In that case, tunneling is always irrelevant for $g<1$, and
the only {\sl relevant}\ coupling corresponds to the scaling field
$V_{1b}$ in Eq.~(\ref{elop}), provided $g<1/2$. For $g>1/2$, the effects
of tunneling ($V_2$) and of the electrostatic coupling $(V_{1b}$) 
can both be treated perturbatively.  We shall therefore focus on the most
interesting strong-interaction region $g < 1/2$ in this subsection,
and omit the irrelevant perturbation $V_{1a}$ as well as tunneling $(V_2)$.
Introducing symmetric and antisymmetric fields \cite{xll}, 
\begin{eqnarray} \label{kanfelder}
  \theta_\pm(x) &=& [ \theta_1(x) \pm \theta_2(x)]/\sqrt2 \; , \\ \nonumber
 \phi_\pm(x) & =&  [\phi_1(x) \pm \phi_2(x)]/\sqrt2 \; ,
\end{eqnarray}
the Hamiltonian decouples,  $H=H_++H_-$, with
\begin{eqnarray} \label{hampm}
  H_\pm &=& \frac12 \int dx [ (\partial_x \phi_\pm)^2 +
(\partial_x \theta_\pm)^2 ] \\ \nonumber &\pm&
(L_c V_{1b}/2) \cos [ \sqrt{8 \pi g}\, \theta_\pm(0) ] \;. 
\end{eqnarray}
An effective coupling strength is defined as
\begin{equation}\label{eff}
 T_B = (c_g/a) [aL_c V_{1b}]^{1/(1-2g)} \;,
\end{equation}
where $c_g$ is a numerical constant of order unity \cite{kf,bistab}.
The boundary conditions (\ref{bc}) also decouple in
 the symmetric/antisymmetric
($r=\pm$) channels.  Effective right- and left-moving
($p=R/L=\pm)$ densities $\bar{\rho}_{pr}(x)$ for these channels
can be defined in analogy to Eq.~(\ref{densbos}). The
new densities again obey the boundary conditions (\ref{bc}), but with 
the effective voltages
\begin{equation} \label{u12}
U_{1,2}\to U_{r=\pm}=(U_1\pm U_2)/\sqrt2 \;.
\end{equation}
It is also useful to 
define the effective current $\bar{I}_r$ in channel $r=\pm$, 
see Eq.~(\ref{currbos}). The current in QW $\alpha=1,2$ is then
given by $I_\alpha=(\bar{I}_+\pm \bar{I}_-)/\sqrt{2}$.

Notably, the full nonlinear
correlated transport problem of crossed LLs therefore completely
decouples into two effective single-impurity problems $r=\pm$ characterized by
effective impurity strength $\pm T_B$, applied voltage $(U_1\pm U_2)/\sqrt2$, 
and interaction strength $2g$.  This single-impurity problem has
been studied in detail, e.g.~by Kane and Fisher \cite{kf}, and the
exact solution for arbitrary interaction strength has been
given in Ref.~\cite{bistab} by combining Eq.~(\ref{bc}) and powerful methods
from boundary conformal field theory.  This solution 
can then be immediately applied to the crossed LL transport problem.
Below we discuss the salient features
for the special case $g=1/4$.  These features 
are characteristic for the strong-interaction regime $g<1/2$.
For weak interactions, $g>1/2$, all inter-wire couplings
may be treated in perturbation theory, and results can be 
found, e.g.~in Ref.~\cite{flensberg}.

\begin{figure}
\begin{center}
\epsfxsize=1.0\columnwidth
\epsffile{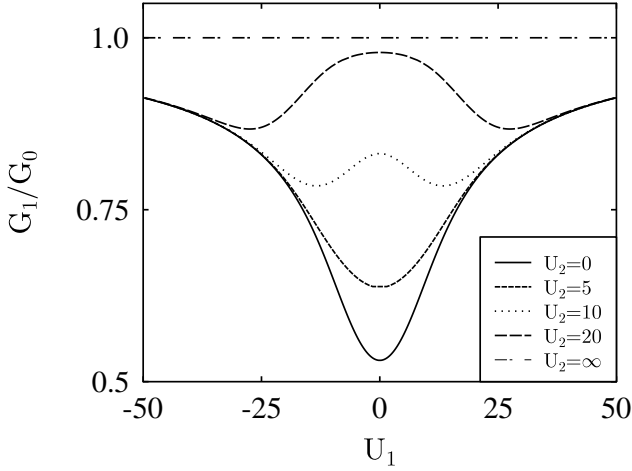}
\caption[]{\label{fig2} 
Nonlinear conductance $G_1(U_1,U_2)=I_1/U_1$
for $g=1/4$, $T=2$ and various $U_2$.
Units are chosen such that $T_B=1$ and $G_0=e^2/h$.}
\end{center}
\end{figure}

The currents through QW $\alpha=1,2$ are 
\begin{equation}\label{g14}
I_\alpha  = (e^2/h)  [U_\alpha - (V_+\pm V_-)/\sqrt2 ] \; ,
\end{equation}
where $V_{\pm}$ obeys the self-consistency relation \cite{bistab}
\begin{equation} \label{g12}
V_\pm = 2 T_B \,{\rm Im}\, \Psi\left(\frac12
+ \frac{T_B + i(U_\pm-V_\pm/2)}{2\pi T} \right) \;,
\end{equation}
with the digamma function $\Psi$. 
Similar but more complicated self-consistency equations
need to be solved for $g\neq 1/4$.  From Eq.~(\ref{g14}), one can verify
that the conductance matrix (\ref{tcond}) fulfills the bounds
\begin{equation} \label{bounds}
 0\leq G_{11}/G_0 \leq 1\;, \qquad -1/2 \leq G_{12}/G_0 \leq 1/2 \;,
\end{equation}
with $G_0=e^2/h$.
The current-voltage relation (\ref{g14}) or, equivalently, the nonlinear
conductances $G_{\alpha \alpha'}$ are very different from
the corresponding results for Fermi liquids \cite{rojo}.
The correlation effects are most
pronounced for $T=0$ \cite{xll}, where perfect zero-bias anomalies,
a strong dependence of $G_{11}$ on the applied cross voltage $U_2$, 
and minima in $G_{11}$ for $|U_1|=|U_2|$ are predicted. 
Such effects are distinct and dramatic signatures of correlations
in a LL, and have found evidence in
recent experiments on crossed multi-wall nanotubes \cite{kim}.
Note that for a Fermi liquid, $G_{11}$ would neither depend on 
$U_1$ nor on $U_2$. 
Thermal fluctuations tend to smear out these phenomena, see  
Fig.~\ref{fig2}, but they remain clearly discernible.

In the linear transport regime, $|U_{1,2}|\ll T$, by
Taylor expanding Eq.~(\ref{g12}) in the small parameter
$(U_1\pm U_2)/T$, we find for $g=1/4$ the voltage-independent
linear conductance,
\begin{equation} \label{gvonT}
 G_1/G_0 =  \frac{1- c \, \Psi'(\frac12+c)}{1+c \, \Psi'(\frac12+c)}
\;, \qquad c=T_B/2\pi T \;.
\end{equation}
More interesting is the {\sl transconductance}, which reads 
to leading order in $(U_1\pm U_2)/T$:
\begin{equation}\label{abv}
G_{12}/G_0  =  \frac{U_1 U_2}{T_B^2} \, c^3
\Psi^{\prime\prime\prime} (1/2 + c) \;.
\end{equation}
Clearly, for very small applied voltages 
(either $U_1$ or $U_2$ approaching zero),
the transconductance vanishes.  As shown in Sec.~\ref{secIIIb},
the vanishing linear transconductance is a
general consequence of the point-like nature assumed for the
contact.  For an extended contact, the linear transconductance need not
vanish.  For low temperatures, $|U_{1,2}|\ll T\ll T_B$, 
Eq.~(\ref{abv}) gives
$G_{12}/G_0  \simeq 2 U_1 U_2/T_B^2$,
while in the high-temperature limit, $T\gg T_B$, we find
$G_{12}/G_0 \simeq \pi U_1 U_2 T_B/ 8T^3$.

Finally, for very strong interactions, $g\leq 0.1$,
bistability effects were reported in Ref.~\cite{bistab}.
As these phenomena require spinless electrons and cannot
be observed in nanotubes, however,
we do not further discuss them here.
Based on the mapping of the transport problem of crossed
LLs to two decoupled single-impurity problems, it is
nevertheless straightforward to obtain the conductance
matrix in closed form in the bistability regime.

\subsection{Short contact}
\label{secIIIb}

Next we turn to a short but nonlocal coupling,
where $L_c$ is of the order of a few lattice spacings $a$.
 For short contact length, 
the bo\-so\-nic fields can be expanded in powers of $x$ around the
center of the coupling region, $x=0$. 
The subsequent RG analysis shows that the only relevant coupling
term is still due to the scaling field $V_{1b}$,
 and one arrives again at the Hamiltonian
(\ref{hampm}). The only difference is the replacement
$V_{1b}\to V_{1b}^\pm$ with 
\begin{equation} \label{intdef}
 L_c V^\pm_{1b} = V_{1b} \int dx  \zeta(x) \cos 
[ 2 (k_{F1} \pm k_{F2} )x ] \; ,
\end{equation}
leading to couplings $T_B^\pm$ as in Eq.~(\ref{eff}).
For large $L_c$,  
$T_B^+\ll T_B^-$ due to the oscillatory integrand.
The resulting physics is similar to the
point-like case of Sec.~\ref{secIIIa}, but in addition exhibits linear
{\sl Coulomb drag} \cite{flensberg,drag}.  Notably, we can
exploit the exact solution
of Ref.~\cite{bistab} for the single-impurity problem
to solve this problem too.

In the remainder of this subsection, we again focus on
the case of $g=1/4$ where the algebra becomes quite simple.
In particular, one has to solve the
self-consistency equations (\ref{g12}) with $T_B\to T_B^\pm$.
For clarity, we consider the special case $U_2=0$, 
and compute  the transconductance $G_{21}$ defined in Eq.~(\ref{tcond}).
Notably, even though $U_2=0$, the 
current $I_2$ and hence $G_{21}$ can be finite (Coulomb drag).
The transconductance follows from
\begin{equation} \label{trrr}
 G_{21}(U_1,T)/G_0 =  \frac{1}{\sqrt{2}} \frac{\partial}{\partial U_1}
( V_- - V_+ ) \;  ,
\end{equation}
where the $V_\pm$ are determined as the $U_2=0$ solutions of Eq.~(\ref{g12})
 with $T_B\to T_B^\pm$.

\begin{figure}
\begin{center}
\epsfxsize=1.0\columnwidth
\epsffile{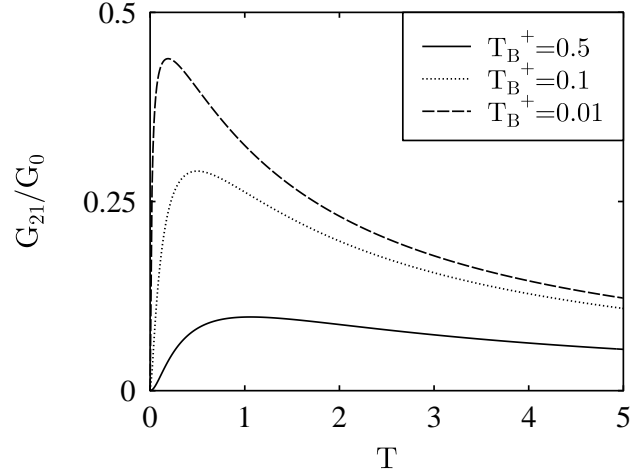}
\caption[]{\label{temps} Temperature dependence of the linear transconductance
for a short contact length at $g=1/4$.  We put $T_B^-=1$. 
Note that $G_{21}\to 0$ for $T\to 0$. 
}
\end{center}
\end{figure}

The {\sl linear transconductance} at $T>0$ can now be finite due to the
nonlocality of the contact \cite{flensberg}. We find
\begin{equation} \label{cover}
 G_{21}/G_0 =   \frac12\sum_{\pm}  \pm
 \frac{1- c_\pm \Psi'(\frac12+c_\pm)}{1+c_\pm\Psi'(\frac12+c_\pm)} \;, 
\end{equation}
where $c_\pm\equiv T_B^\pm /2\pi T$.
The temperature dependence of the $g=1/4$ linear transconductance (\ref{cover})
is shown in Fig.~\ref{temps}.
We observe that for a 
long contact, $T_B^+\ll T_B^-$, the
perfectly quantized transconductance $e^2/2 h$ is approached at
low but finite temperature. This is the 
{\sl absolute drag effect}\ reported by Nazarov and 
Averin \cite{drag} for a long extended contact. 
Equation (\ref{cover}) then describes how the absolute drag effect 
is thermally destroyed at high temperatures.

However, as $T\to 0$, the linear transconductance {\sl vanishes}
throughout the strong interaction regime $g<1/2$.
(This is an artefact of the locality assumption
and will not be true for extended contacts, see below.)
In the zero-temperature limit, one can obtain the
nonlinear $g=1/4$ transconductance for $|U_1|\ll T_B^\pm$ and $U_2=0$ 
from Eqs.~(\ref{trrr}) and (\ref{g12}),
\begin{equation}
G_{21}/G_0 = (U_1/4)^2 [ (T_B^+)^{-2} - (T_B^-)^{-2} ] + O([U_1/T_B^+]^4)\;.
\end{equation}


Finally, let us address the role of a difference in the Fermi momenta, 
$\delta k_F=k_{F1}-k_{F2}$, which could experimentally
be tuned by varying the mean chemical potential in
one QW relative to the other.
In the absence of electron tunneling between the QWs,
the nonlinear conductance matrix $G_{\alpha\alpha'}(U_1,U_2)$ 
for a short contact does only depend on $\delta k_F$
via the $T_B^\pm$ following from Eq.~(\ref{intdef}).
In fact, for a strictly local contact ($T_B^\pm=T_B$), there
is no dependence on $\delta k_F$ at all unless there is
tunneling.   As discussed in Sec.~\ref{secIV}, this
effect can be used to experimentally disentangle the effects of tunneling
and electrostatic coupling.

\subsection{Arbitrary contact length}
\label{secIIIc}

Next we study an arbitrarily long contact, where $L_c$ can
even approach the system length $L$.
Here tunneling can be a relevant
perturbation under the RG transformation, provided the interaction is not
too strong.   Following the analysis of Refs.~\cite{kusm,yakov},
for $g>g'=\sqrt{2}-1\simeq 0.414$, tunneling is relevant.
In fact, for $g>g^{\prime\prime}=1/\sqrt{3}\simeq 0.577$, 
tunneling $(V_2)$ is even more relevant than the electrostatic
coupling $V_{1b}$.  In this regime, one should then first treat
the tunneling.  Below we focus on the strong-interaction case,
$g<g^{\prime\prime}$. 
For $g' < g< g^{\prime\prime}$, the main effect of tunneling
is a renormalization of the electrostatic coupling \cite{yakov}.
In the following we assume that $V_{1b}$ contains this renormalization
and then omit the scaling field $V_2$.  
Since for a short contact, tunneling is always irrelevant, 
such a reasoning can be applied for arbitrary $L_c$ in the
regime $g<g^{\prime\prime}$, 
in particular for the case $g=1/2$ investigated below.

Taking into account the $V_{1a}$ operator in the bulk (i.e., for
$L_c=L$) causes a renormalization of the
interaction constants, thereby splitting them \cite{klesse},
\begin{equation} \label{kl}
g\rightarrow g_{\pm}=\frac{g}{\sqrt{1\pm g V_{1a}}} \; .
\end{equation}
For $L_c\ll L$, however, the scaling field $V_{1a}$ is {\sl irrelevant}. 
Below we shall neglect the weak splitting (\ref{kl}). 
Following our analysis, this
could only create a problem for very long contacts, $L_c\approx L$.
The case $g=1/2$ then permits a full solution of this transport
problem for arbitrary $L_c$. 
The resulting effective Hamiltonian is
\begin{eqnarray}  \nonumber
 H = H_0 &+& V_{1b} \int d x  \zeta(x)  \sin \left[
2 k_{F1} x + \sqrt{2 \pi} \,\theta_1(x) \right] \\ \label{V1Termx}
&\times& \sin \left[- 2 k_{F2} x + \sqrt{2 \pi} \,\theta_2(-x) \right] \; ,
\end{eqnarray} 
where the spatial coordinate along QW $\alpha=2$ has been changed
to $-x$. As we shall see later, this mirroring of the axis is crucial to 
ensure correct anticommutation relations between new fermionic fields. 
Expanding the product of $\sin$ terms,
$\cos$ functions of the sum and the
difference of the fields $\theta_\alpha$ emerge.
In Ref.~\cite{drag}, only the difference term was kept,
since the other term is highly oscillatory for long contacts
and does not contribute for $L_c\gg a$. 
However, a proper description of the crossover from a short to
a long contact requires to consider the full coupling (\ref{V1Termx}).
Below we take Eq.~(\ref{contact}) for $\zeta(x)$.

\subsubsection{Refermionization}

Remarkably, the transport problem posed by Eqs.~(\ref{V1Termx}) and
(\ref{bc}) can be solved exactly for
arbitrary $L_c$ by refermionization.
Switching to the chiral fields, 
\begin{equation} \label{chiral}
 \varphi^\alpha_{R,L}= \sqrt{\pi}(\phi_\alpha\pm \theta_\alpha) \; ,
\end{equation}
we obtain with $\delta k_F = k_{F1} - k_{F2}$ and $2k_F=k_{F1}+
k_{F2}$
\begin{eqnarray*} 
 H &=& \frac{1}{8\pi} \sum_{\alpha=1,2} \int dx \left[ (\partial_x
 \varphi^\alpha_L)^2 + (\partial_x \varphi^\alpha_R)^2 \right] 
 \\  &-& V_{1b} \int dx \zeta(x) \Bigg\{ 
 \sum_{p=\pm} e^{ip 2 \delta k_F x}
e^{i p (\varphi^1_R(x)-\varphi^2_L(-x))/\sqrt{2}}  \\ 
 &\times& e^{-i p (\varphi^1_L(x)-\varphi^2_R(-x))/\sqrt{2}} -  
 e^{i p [4k_F x+ (\varphi^1_R(x)+\varphi^2_L(-x))/\sqrt{2}]} 
  \\ &\times& e^{-i p (\varphi^1_L(x)+\varphi^2_R(-x))/\sqrt2}  
\Bigg\} \; .
\end{eqnarray*}
We then employ the slightly modified refermionization 
transformation of Luther and Emery \cite{lutheremery}, 
\begin{eqnarray} \label{novyepolya}
 \Psi_{1,2}(x) &=& \frac{\eta_{1,2}}{\sqrt{2 \pi a}}\, e^{ i2 k_{F1}x + i(\varphi^1_R(x)\mp
\varphi^2_L(-x))/\sqrt{2}} \; , \\ \nonumber
 \Psi_{3,4}(x) &=& \frac{\eta_{3,4}}{\sqrt{2 \pi a}}\, e^{\pm i 2 k_{F2}x \mp
i (\varphi^2_R(-x)\mp \varphi^1_L(x))/\sqrt{2}} \; .
\end{eqnarray}
Special Majorana fermions $\eta_i$ ensure the correct anticommutation
relations between new operators, see Ref.~\cite{lltube2}. However, all
chemical potentials of the new particles can be incorporated into the 
refermionized Hamiltonian ($\lambda=2\pi a V_{1b}$)
\begin{eqnarray}
 H &=& - \int dx \Big\{\sum_{j=1,2} \Psi_j^\dag(x) \, (i \partial_x + 2
k_{F1})  
\nonumber
 \Psi^{}_j(x) 
\nonumber \\
&-&  \sum_{j=3,4} \Psi_j^\dag(x) \, (i \partial_x \mp 2 k_{F2}) 
\nonumber
 \Psi^{}_j(x) \Big\} \\ 
 &+&  i \lambda \int dx  \zeta(x)\Big \{
\Psi_1^{\dag}(x) \Psi^{}_3(x) \label{refhams}
+ \Psi^{}_1(x) \Psi^{\dag}_3(x) 
 \\ &-& \Psi^{}_2(x)
\Psi^\dag_4(x) - \Psi^\dag_2(x) \Psi^{}_4(x) \Big\} \; . \nonumber
\end{eqnarray}
Then they disappear in the definition (\ref{novyepolya}). 
This problem permits an exact solution, since we are left
with a quadratic Hamiltonian expressed in terms of the $\Psi_i(x)$
operators.  We obtain this solution via the equations of motion 
\begin{eqnarray} \label{eqm1}
(\partial_t \pm \partial_x) \Psi_{1,3}^\dag(x,t) &=& \pm\lambda \zeta(x)
\Psi_{3,1}^{\dag}(x,t)  \\ \label{eqm2}
(\partial_t \pm \partial_x) \Psi_{2,4}^\dag(x,t) &=& \mp\lambda \zeta(x)
 \Psi_{4,2}^{\dag}(x,t)  \;.
\end{eqnarray}
They describe free chiral fermions with linear dispersion  
outside of the contact area.  
The contact region, $|x|\leq L_c/2$, then acts as a scatterer. 
Denoting by $a_i^\dag(k)$ ($b_i^\dag(k)$)
the momentum-space creation operator for
fermions moving towards (away from) the scatterer,
their respective particle densities
can be related by a transmission matrix $D_{ij}(k)$,
\begin{equation}
 \langle: b^\dag_{j}(k) b^{ }_j(k) :\rangle = \sum_{i=1}^4 D_{ji}(k)\,\langle:
 a^\dag_{i}(k) a^{ }_i(k) :\rangle \; .
\end{equation}
The matrix $D_{ij}$ can be found in closed form from Eqs.~(\ref{eqm1})
and (\ref{eqm2}), see Sec.~\ref{trans}.

To solve the full transport problem, we also need to re-express the
boundary conditions (\ref{bc}) in the new fermion basis.
Defining $\rho_i(x)=\langle : \Psi_i^\dag(x) \Psi_i(x) : \rangle$ as
density of the new fermions, for $g=1/2$,
we get the relation to the previous  densities $\rho_{p\alpha}(x)$ (where
$p=\pm=R/L$): 
\begin{eqnarray*}
 \rho_{p1}(x) &=& ( 1+2p) [\rho_1(x) + \rho_2(x) ]
+ (1 -2p) [ \rho_3(x) - \rho_4(x) ] \;, \\
 \rho_{p2}(-x) &=& (2p- 1) [\rho_1(x) - \rho_2(x) ]
 - (1+2p)[\rho_3(x) + \rho_4(x)] \; .  
\end{eqnarray*}
The normal ordering should be performed with respect to the ground state, 
given by a Fermi sea filled up to $k=2 k_{F1}$ and $k=\mp 2 k_{F2}$ for 
channels (1,2) and (3,4), respectively.
Plugging these last relations into Eq.~(\ref{bc}) gives the
boundary conditions: 
\begin{eqnarray} \label{bc1}
 3 \rho^-_1+ 3\rho^-_2 +\rho^-_3-\rho^-_4 &=& U_1/4 \pi \;, \\ 
\nonumber
 -\rho^+_1+ \rho^+_2 -3\rho^+_3-3\rho^+_4&=& -U_2/4 \pi \;, \\
\nonumber
 \rho^+_1+  \rho^+_2 +3\rho^+_3-3\rho^+_4 &=& -U_1/4 \pi\;, \\
\nonumber
 -3 \rho^-_1+ 3\rho^-_2 -\rho^-_3-\rho^-_4 &=& U_2/4 \pi \; ,
\end{eqnarray}  
where $\rho^{\pm}_i=\rho_i(\pm L/2)$. 
For the current, one gets 
\begin{equation} \label{tokI}
 I_1 = 2 [ \rho_1(x) + \rho_2(x) - \rho_3(x) + \rho_4(x) ] \; .
\end{equation}
Here $x$ is arbitrary due to the continuity equation.
The incoming free fermions must obey the Fermi distribution,
\[
 \langle: a^\dag_{j}(k)\,a^{ }_j(k) :\rangle = n_F(k-\mu_j) - n_F(k) \; ,
\]
with $n_F(E)=1/[\exp(E/T)+1]$. Therefore we obtain
\begin{equation} \label{plota}
 \rho_j^\mp = \int \frac{d k}{2 \pi} \langle: a^\dag_{j}(k)\,a^{ }_j(k)
:\rangle \equiv \mu_j \; , 
\end{equation}
where $-$ sign should be taken for channels $1$ and $2$, and $+$ for $3$
and $4$. The effective chemical potentials $\mu_j$ have to be 
computed self-consistently, see below.
The outgoing densities are then given by
\begin{equation} \label{plotb}
 \rho_j^\pm =  \sum_{i=1}^4
 \int \frac{d k}{2 \pi} D_{ji}(k) [ n_F(k-\mu_i) - n_F(k)] \; ,
\end{equation}
where $\pm$ apply to the channels (1,2) and (3,4), respectively.

\subsubsection{Transmission matrix}
\label{trans}
The transmission matrix contains only one
independent element. Obviously, $D_{13}=D_{31}$, $D_{11}=D_{33}$, 
$D_{24}=D_{42}$, and $D_{22}=D_{44}$ because of the system symmetry.  
All other matrix elements vanish since the channels (1,3) and (2,4) 
fully decouple.
Furthermore, since the Hamiltonian (\ref{refhams}) conserves the net particle
numbers in channels (1,3) and (2,4), one has the additional relations
$D_{13}(k)= 1-D_{11}(k)$ and $D_{24}(k)=1-D_{22}(k)$.
The last simplification stems from the symmetry of the
equations of motion and reads $D(k)=D_{22}(k)=D_{11}(k)$.  

Let us now explain how to find $D(k)$. Since a
right-mover in channel 1 is scattered to a left-mover in channel 3 within the 
contact, we 
can regard them as a single species which is backscattered by the contact.  
Then the problem reduces to the determination of the penetration coefficient 
of 1D fermions through a rectangular barrier. The corresponding result can be
found, e.g. in Ref.~\cite{LLIII}, 
\begin{eqnarray}
 D(k)= \frac{4 k^2 |\chi|^2}{\lambda^4 |\sin[L_c \chi]|^2 + 4 k^2
 |\chi|^2} \; ,
\end{eqnarray}
where $\chi^2=k^2-\lambda^2$. Next we insert Eqs.~(\ref{plota}) and 
(\ref{plotb}) into the boundary conditions (\ref{bc1})
and compute the current from Eq.~(\ref{tokI}). 

\subsubsection{Linear transconductance}
The {\sl linear transconductance} $G_{12}$
can be found in closed form, where we focus on $T=0$
and $\delta k_F=0$ again.  Notably, this quantity 
does not vanish in general as it would for a point-like coupling.
Linearizing Eq.~(\ref{plotb}) in the chemical potentials $\mu_i$,
we need to solve the now linear system of equations (\ref{bc1}) for
the $\mu_i$, yielding
\begin{eqnarray} \label{lfinL}
2G_{12}/G_0 = \frac{1- D(2 k_F)}{1- D(2 k_F)/2} \; ,
 \end{eqnarray}
For uncoupled nanotubes, $\lambda\to 0$, we find 
$D(2k_F)=1$ and thus $G_{12}=0$.
On the other hand, for strongly coupled tubes, $\lambda\to \infty$, 
we get $D(2 k_F)=0$ and hence recover the absolute Coulomb drag, 
$G_{12}=e^2/2h$, in this limit.
For small contact lengths, it is possible to perform a Taylor expansion of 
$D(k)$. The resulting linear transconductance behaves according to 
\begin{equation}
 2G_{12}/G_0 = \frac{1}{8} (\lambda L_c)^2 \left(\lambda /k_F\right)^2 \; .
\end{equation}
The local approach of Section \ref{secIIIb} for $g=1/2$  gives
\begin{equation}
  2G_{12}/G_0 = \frac{1}{3} (2 \pi V_{1b} L_c)^2 \left( k_F L_c \right)^2 \;
  , 
\end{equation}
where $V_{1b}$ denotes the coupling strength of the related
local problem, see Eq.~(\ref{intdef}). These two results suggest the following
relation between the coupling constants 
\begin{equation}
 V_{1b} = \frac{1}{4 \pi a} \left( \frac{3}{2} \right)^{1/2} 
 \left( \frac{\lambda}{k_F} \right)^2 \frac{1}{L_c} \; ,
\end{equation}
which establishes a connection between the local approximation and the exact 
solution for the system with a short contact. The corresponding amplitudes
$V_{1b}^\pm$ are then given by a Taylor expansion of Eqs.~(\ref{intdef}) to
second order in $L_c/a$, implying that $V_{1b}^-= V_{1b}$ and 
$V_{1b}^+=V_{1b}(1-2(k_F L_c)^2/3)$.  

Moreover, employing the Sommerfeld expansion, the finite-temperature 
corrections   
\[
\Delta
 G_{12}(T) = G_{12}(T)-G_{12}(T=0)
\] 
to Eq.~(\ref{lfinL}) are of the form
\[
\Delta G_{12}(T)/G_0 = -\frac{\pi^2 T^2}{6} \frac{\partial^2
D(k=2 k_F)}{\partial k^2} 
\frac{1}{(2-D(2 k_F))^2}
\]
\begin{figure}
\epsfxsize=1.0\columnwidth
\epsffile{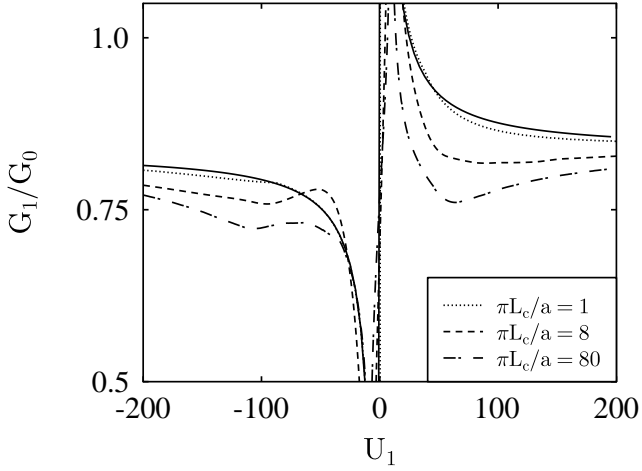}
\caption[]{\label{fig5b} 
Nonlinear conductance $G_{1}$ (in units of $G_0=e^2/h$)
as a function of $U_1$ for $g=1/2, T=0, 
\lambda=4,  U_2=25$, and various contact lengths. 
The solid curve is obtained analytically for the shortest contact
using the point-like coupling (\ref{hampm}).}
\end{figure}

\label{secnonlin}

In the general nonlinear case, the current has to be evaluated numerically,
see Fig.~\ref{fig5b}. All curves show a singularity
at $U_1\to 0$, independent of $L_c$. This  singularity implies 
non-vanishing current for zero applied voltage and reflects Cou\-lomb drag.

\subsubsection{Absolute Coulomb drag}

The absolute Coulomb drag reported in Refs.~\cite{flensberg,drag} can be 
recovered by our $g=1/2$  solution. For a very long contact ($L_c=L$),
$H$  can be diagonalized by means
of a Bogoliubov transformation. 
Consider, e.g.~channels $j=1$ and 3,  
with 
\[
 \Psi_j(x) = \int \frac{d k}{2 \pi} e^{i k x} c_j(k) \; .
\]
Allowing for $\delta k_F\neq 0$ and switching to
\begin{eqnarray*}
 \alpha_1(k) &=&  c_1(k) \cos \delta_k + i
c_3(k) \sin \delta_k
 \;, \\
 \alpha_3(k) &=& - c_1(k) \sin \delta_k + i c_3 (k) \cos \delta_k\;  ,
\end{eqnarray*}
where $2\delta_k =  \arctan (\lambda/(k+\delta k_F))$,
we find for the channels 1 and 3:
\begin{eqnarray*}
 H_{1,3} &=& \int \frac{d k}{2 \pi}  \left[2 k_F + \sqrt{(k+ \delta k_F)^2 +
 \lambda^2} \right] \, \alpha_1^\dag(k) \alpha^{}_1(k)  \\
 &+& \left[2 k_F - \sqrt{(k+ \delta k_F)^2 +
 \lambda^2} \right] \alpha_3^\dag(k) \alpha^{}_3(k)  \; . 
\end{eqnarray*}
The Hamiltonian $H_{2,4}$ follows accordingly
by replacing $\delta k_F \leftrightarrow 2k_F$.

We can now distinguish two situations: (i) $|\delta k_F| \ll \lambda$, 
where the  channels $j=1,3$ are gapped and therefore
transport for $U_{1,2} \ll
\lambda$ is strongly suppressed, and (ii) $|\delta k_F| \ge \lambda$,
with no gap. Since $k_F \gg \lambda$, transport properties
involving  the channels 2 and 4 will always be perfect.    
For case (ii), the coupling $\lambda$  will thus not significantly
affect transport, i.~e. up to weak perturbations, the quantized ideal
currents are found.  However, 
for case (i), we obtain at $T=0$ 
\begin{equation} 
I_1 = (e^2/2 h) (U_1+U_2) \;,
\end{equation}
just as is expected for absolute Coulomb drag.
This simple calculation shows that sufficiently large 
$\delta k_F$ can destroy the perfect Coulomb drag.  Experimentally,
this could be achieved by uniformly shifting the chemical potentials
of the two reservoirs attached to the same wire.

\section{Weakly coupled reservoirs}
\label{secX}

How important is the requirement of good (adiabatic) contact
to the reservoirs assumed in our theory so far?
Is it still possible to observe the characteristic
effects of crossed LL transport and Coulomb drag
with weakly coupled reservoirs?  To study this point,
we assume now that reservoirs are coupled to the QWs by tunnel
junctions with (identical) dimensionless tunnel conductance  $T_0\ll 1$.
In this section, we consider only the limit of zero temperature 
and sufficiently large applied voltage such that transport proceeds
incoherently at all contacts.

The applied voltage $U_1$ will then split up into three
voltage drops: First, the part $U_1^\prime$ drops at each 
junction, and $U_1^{\prime \prime}$ at the contact between
the nanotubes, where $U_1=2U_1^\prime+U_1^{\prime\prime}$.  The current 
$I_1$ injected into the QW from the (say, left) reservoir is then
\begin{equation} \label{new1}
 I_1 = T_0 G_0  U_1'  |a U_1'|^{\nu_g-1} \; .
\end{equation}
Here the exponent $\nu_g=1/g$ applies to the case
of an end-contacted (spinless single-channel) QW, while
$\nu_g=(g+1/g)/2$ for a bulk-contacted QW \cite{lltube2}.
If the voltage drop $U_1^{\prime\prime}$ 
across the crossing point is significantly higher than $U_1'$, 
the correlation effects described in the previous section dominate the
transport process, and we may use the strong-coupling form of
the current  \cite{xll}, 
\begin{equation} \label{neww}
 I_1= G_0 T_B \sum_{\pm} \mbox{sgn}(U_1^{\prime\prime}
 \pm U_2^{\prime\prime}) 
(|U_1^{\prime\prime}\pm U_2^{\prime\prime}|/T_B)^{1/g-1} \;.
\end{equation}
Since both Eq.~(\ref{new1}) and (\ref{neww}) must give the same
current, the condition $U_1^{\prime\prime} \gg U_1^\prime$ leads to 
the estimate
\begin{equation} \label{us}
U_1, U_2 \gg  T_B
 [T_0(aT_B)^{\nu_g-1}]^{-1/(\nu_g+1-1/g)}  \;.
\end{equation}
Once this condition is satisfied, typical crossed LL
effects can be observed even for
non-adiabatic (weak) coupling to the voltage reservoirs.   
Evidently, Eq.~(\ref{us}) cannot be satisfied for end-contacted
QWs. This is because the power-law exponent
in Eq.~(\ref{us}) is always negative (namely $-1$), 
and with $T_0\ll 1$ the inequality cannot be satisfied  
except for unreasonably high $U_1$.
However, assuming that we have {\sl bulk-contacted QWs}
with $g< \sqrt{2}-1 \simeq
0.414$,  the power-law exponent changes sign, 
and then the condition (\ref{us}) can be fulfilled even for
very small applied voltage $U_1$.

\section{Electron tunneling}
\label{secIV}

Let us now address the influence of electron tunneling between the QWs.
For clarity, we shall focus on the case of a strict 
point-like contact $(L_c\to 0)$ as in Sec.~\ref{secIIIa}.
Assuming $g<1$, tunneling is then irrelevant under the RG and can
be treated to lowest order in the tunneling matrix element $t$ 
(unless $t$ is very large).  In addition, we restrict ourselves
to small applied voltages, $|U_{1,2}|, \delta k_F \ll T_B$, 
at zero temperature, where the tunneling density of states (TDOS) 
describing electron tunneling into
QW $\alpha=1,2$ (at $x=0$) 
{\sl in the presence of the electrostatic coupling} carries
the standard power-law suppression factor \cite{kf}
\begin{equation}\label{tdos}
\rho(E) \propto \Theta(E) E^{1/g-1} \;.
\end{equation}
Of particular interest will be the effect of $\delta k_F=k_{F1}-k_{F2}$
on the transport properties in the presence of tunneling
(without loss of generality, we consider $\delta k_F > 0$).

Tunneling  modifies the currents flowing to the left and to the
right of the coupling point $x=0$.  Under a golden rule
calculation, these currents can be easily found from the
rates for tunneling of a $p=\pm=R/L$ mover in QW $1$ into
a $p'$ mover in wire 2:
\begin{equation}\label{tunn}
 \Gamma_{p1 \to p'2} = |t|^2 \Theta(\Delta_{pp'})
 \int_0^{\Delta_{pp'}} \, dE  \, \rho(E) 
\rho(\Delta_{pp'} - E) \; .
\end{equation} 
Because of the pointlike coupling, there is no momentum conservation,
and the tunneling electron can change its chirality.
In Eq.~(\ref{tunn}),  
\begin{equation}
\Delta_{pp'}= \delta k_F/g + \mu_{p1}-\mu_{p'2} \;,
\end{equation}
where $\mu_{p\alpha}$ is the chemical potential of a $p$ mover in QW
$\alpha=1,2$ (relative to the respective mean chemical potential),
 which is determined by the applied voltages $U_{1,2}$
and the electrostatic coupling $T_B$.  
Simple dimensional scaling gives from Eq.~(\ref{tunn})
\begin{equation} \label{tunn2}
\Gamma_{p1\to p'2} \propto \Delta_{pp'}^{2/g-1} \;.
\end{equation}
For $\delta k_F \gg g |U_1,U_2|$, the tunneling rates
become independent of the applied voltages $U_{1,2}$.
In this limit, tunneling rates are large, and therefore 
tunneling can potentially modify the transport behaviors 
discussed in Sec.~\ref{secIII}.  However, for $\delta k_F \ll
g |U_1,U_2|$, the rates are power-law suppressed, and hence
tunneling rates become very small.  In fact, since the
power-law exponent is larger than the one 
in the absence of tunneling, see Eq.~(\ref{neww}),
we expect that tunneling
leads only  to {\sl subleading} corrections for small $\delta k_F$.
The role of tunneling in a concrete experiment could then
be revealed by simply tuning $\delta k_F$ from zero to 
large values.  If this leads to a destruction of the crossed
LL scenario, tunneling matrix elements must be sizeable
and should be taken into account for large $\delta k_F$.

\section{Conclusions and applications}
\label{secV}

Let us briefly summarize the conclusions and major findings
of our paper.  We have studied transport through
two quantum wires or carbon nanotubes that are coupled
along a contact length $L_c$. Assuming that the wires are
in the Luttinger liquid state with sufficiently strong 
interactions,  the main coupling mechanism is of electrostatic
nature, and tunneling provides only perturbative corrections
to the problem.  The electrostatic coupling leads to qualitatively
new nonequilibrium transport behaviors compared to the case of
Fermi liquid wires, namely crossed Luttinger liquid effects and
Coulomb drag.  Crossed Luttinger liquid effects are characterized
by pronounced zero bias anomalies, dependencies of transport
currents on applied cross voltages, and resonant behaviors
at finite voltages.  For extended contacts, very pronounced
Coulomb drag effects arising at low temperatures were found.
For short contacts, the linear transconductance has a maximum at
a finite temperature and approaches zero as $T\to 0$. 

For sufficiently long contacts
Coulomb
drag can even be perfect at low temperatures,
with the transconductance approaching its largest
possible value $e^2/2h$.  We have presented an exact
solution of the full transport problem valid for arbitrary
contact length $L_c$ at the special interaction strength $g=1/2$. 
In addition, 
the nonlinear properties of Coulomb drag and their relation to 
crossed Luttinger liquid effects were clarified. 
Our study has assumed adiabatic coupling to the voltage reservoirs.
However, as outlined in Sec.~\ref{secX}, for sufficiently
strong interactions, the same characteristic effects 
are expected for bulk-contacted nanotubes, where the nanotubes are
in weak (tunneling) contact with the leads.  
In addition, we have shown that the tunneling between the
nanotubes should cause only subleading corrections to 
the behaviors outlined here for small $\delta k_F=k_{F1}-k_{F2}$.
However, by varying this quantity, the role of tunneling can
be easily determined in practice.  For large $\delta k_F$, 
the characteristic correlation effects will be washed out
by a sufficiently large tunneling matrix element. 

\begin{figure}
\begin{center}
\epsfxsize=0.5\columnwidth
\epsffile{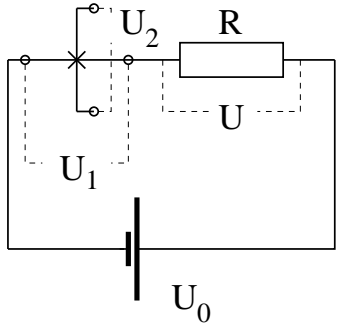}
\caption[]{\label{fig7} 
Amplification circuit based on crossed nanotubes. For fixed $R$ and $U_0$ it
is possible to achieve $|\partial U_2/\partial U_1| \gg  1$. 
}
\end{center}
\end{figure}

\begin{figure}
\begin{center}
\epsfxsize=0.85\columnwidth
\epsffile{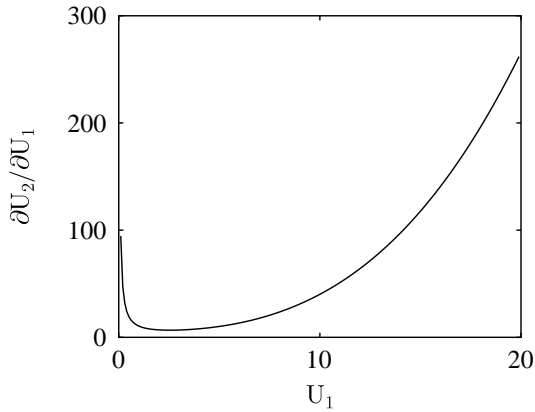}
\caption[]{\label{fig7a} 
Voltage amplification ratio for $R G_0=1$ at $T=0$ and $g=1/4$ (here $T_B=1$).
}
\end{center}
\end{figure}

\begin{figure}
\begin{center}
\epsfxsize=0.65\columnwidth
\epsffile{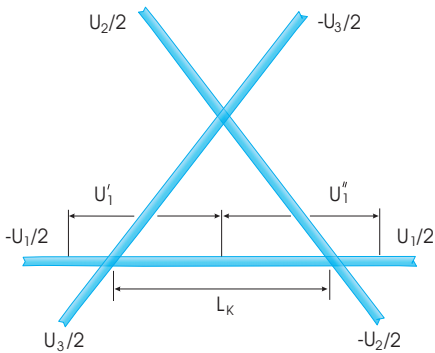}
\caption[]{\label{fig8} 
Triangular setup of nanotubes coupled by  (local)
electrostatic interactions.
}
\end{center}
\end{figure}

Let us conclude the paper with two possible applications.
Crossed nanotubes could be used for 
{\sl voltage amplification}, e.g.~by using the circuit shown in
Fig.~\ref{fig7}. The circuit equation 
 $U_1[1+R G(U_1,U_2)]=U_0$ 
can be easily solved for $g=1/4$ and $T=0$, with the results depicted 
in Fig.~\ref{fig7a}.
Clearly, the voltage amplification ratio
$\partial U_2/\partial U_1$ can be tuned to  extremely
high values.
A more complex setup built up of three nanotubes
coupled to each other in a star-like manner is shown 
in Fig.~\ref{fig8}. 
If the distance $L_K$ between the contact points exceeds
the length scales $v/T$ or $v/U_i$, transport proceeds in
an incoherent way
and can thus be modelled as a sequence of crossed LLs.
The voltage drops $U_i'$ and $U_i^{''}$ arising at the respective contacts
obey $U_i=U_i'+U_i^{''}$, and now effectively
play the role of two-terminal voltages. 
This allows for the direct application of the results of
Sec.~\ref{secIII}, and the whole current-voltage
characteristics of such a setup can be obtained. 
We only point to one interesting consequence of this
solution, namely the possibility of a 
{\sl voltage measurement}\ involving only electrostatic contact.
Setting, say, $U_3=0$, the current $I_1$
is seen to vanish once $U_1=-\kappa U_2$, 
where $\kappa$ is constant and assumed to be known from
previous measurements.  The (presumably unknown) voltage 
$U_2$ can thus be determined by tuning $U_1$ to the point where $I_1=0$.

\begin{acknowledgement}
We thank A.~O.~Gogolin and H.~Grabert for discussions, and
the Deutsche For\-schungs\-ge\-meinschaft for support
under Grant No.~GR 638/19-1 and the Gerhard-Hess program.
\end{acknowledgement}

\end{document}